\documentclass[12pt,a4paper]{article}
\usepackage{amssymb,amsfonts,amsmath}
\usepackage{color}
\usepackage{epsfig, euscript, graphics}

\begin{document}

\title{Two faced Janus of quantum nonlocality}

\author{Andrei Khrennikov\\ 
Linnaeus University, International Center for Mathematical Modeling\\  in Physics and Cognitive Sciences
 V\"axj\"o, SE-351 95, Sweden}

\maketitle

\abstract{This paper is a new step towards getting rid of nonlocality from quantum physics. This is an attempt to structure 
the nonlocality mess. ``Quantum nonlocality'' is Janus faced. One its face 
is projection (Einstein-L\"uders) nonlocality and another Bell  nonlocality. 
The first one is genuine quantum nonlocality, the second one is subquantum nonlocality.
Recently it was shown that Bell ``nonlocality'' is a simple consequence  of the complementarity principle. We now show that projection  nonlocality has no  connection with physical space. Projection state update is generalization  of the well known 
operation of probability update used in classical  inference.  We elevate the role of interpretations of a quantum state. By using the individual (physical) interpretation, one can really get the illusion of a spooky action at a distance resulting from L\"uders' state update. The statistical interpretation combined with treating the quantum formalism as machinery for update of probability is known as the V\"axj\"o interpretation. Here one   follows the standard scheme of  probability update adjusted to the quantum calculus of probability. The latter is based on operating with states represented by vectors (or density operators). We present in parallel classical and quantum probability updates. From this presentation, it is clear that both classical and quantum  ``faster-than-light change of statistical correlation'' take place in  mental and not physical space.}

\section{Introduction}

As was emphasized in recent paper \cite{NL} (see also \cite{ABELL}), the notion ``quantum nonlocality'' is really misleading. One of the difficulties  in struggling with nonlocality is that (as was pointed in \cite{ARV}) the present situation  is the real mess. Surprisingly, this mess-problem is ignored and it is commonly claimed that quantum physics is ``nonlocal'' (without specifying what this means  concretely). 

Personally, I got the first signal that quantum nonlocality is Janus faced from the talk 
of A. Aspect at one of the V\"axj\"o conferences (see also his papers \cite{AA0,AA1}). He started his talk not from the Bell inequality \cite{Bell1,Bell2} and its violation (as could be expected), but with the projection postulate (in L\"uders' form \cite{Luders}\footnote{The projection postulate is often associated with von Neumann. But, von Neumann \cite{VN} suggested to use it only for observables with non-degenerate spectra. And in the Bell framework, observables 
have degenerate spectra. For such observables, von Neumann proposed to use the machinery which later was developed to theory of quantum 
instruments \cite{DV,Oz,Oz1}.}  and its nonlocal consequences.  He pointed that this {\it  projection nonlocality} 
is really counter-intuitive and that, to find a proper physical picture, one has to introduce hidden variables. From this viewpoint, quantum theory really cry for hidden variables! However, on this way, as is commonly accepted, one confronts with
the Bell inequality and proceeds towards {\it Bell nonlocality} - nonlocality of models with hidden variables.
Aspect cleaned the well-known EPR-reasoning \cite{EPR} from the elements of reality and statements with probability one.  
His presentation is essentially clearer than the original EPR-argument.

In paper \cite{NL}, I tried to destruct this huge plant - Bell nonlocality - which grew up from the projection-seed.
The main message from \cite{NL} is that violation of the Bell type inequalities is the
straightforward consequence of {\it local incompatibility of observables}. Thus the Bell test 
can be interpreted as a very special test of the Bohr's complementarity principle. Purely quantum 
treatment of the Bell type inequalities does not leave any place for nonlocal speculations. (Similar conclusion can be found 
in works of De Muynck \cite{Muynck}, Boughn\cite{Boughn1},  Griffiths \cite{Griffiths})\footnote{At the same time, some subquantum models can have nonlocal features, e.g., 
Bohmian mechanics. But, this problem has not so much to do with quantum theory. These models are just fruits of human imagination and in principle  they do not have anything to do with physics (may be besides Bohmian mechanics). As was pointed by Bell, the emphasis of such models may be just the sign of lack of imagination.} The aim of the present paper is to destroy even the seed - projection 
nonlocality. 

Projection nonlocality will be analyzed in very detail. We show that it is fictitious; this is ``nonlocality'' 
of probability update; it is identical with   ``nonlocality'' of classical probability update. 

We also emphasize the role of interpretations of a quantum state, individual (physical) vs. statistical (and its special version, {\it the V\"axj\"o interpretation} \cite{V2}). 

In fact, I started writing this paper as a reply to the following comment of Johan Summhammer on my previous article  \cite{NL}: 

\medskip

{\it ``I have looked at your ``Get rid of non locality...'' and I agree at the formal level, that your are only dealing with incompatible operators. But the empirical fact remains, that setting a measurement operator on site A has an instantaneous ``influence'' on site B, if you later calculate conditional statistics between the data of site A and site B. And this faster-than-light change of statistical correlation is already empirically proven. Sure, no transfer of information. But more than ``no influence''. Which name would you assign to this phenomenon?''}

\medskip

I shall try to answer to Summhammer's question in  section \ref{SRT}.

This paper was also stimulated by the recent work of A. Plotnitsky \cite{PARX} (in turn stimulated by \cite{NL,ARV}) who analyzed quantum nonlocality in the framework of the original EPR presentation and the debate between Einstein and Bohr \cite{EPR,BR}. He operated with the notion of Einstein nonlocality which is similar to the notion of  projection nonlocality. In any event, Einstein nonlocality is rooted in the projection postulate.  Plotnitsky  concluded that  spooky action at a distance  is in fact  {\it ``spooky predictions at a distance''.} His suggestion is the important step towards modification of the present misleading terminology. However, I am neither happy with Plotnitsky's terminology, concretely with ``spookiness'' of predictions 
(see section \ref{SPD}).       

I want to elevate the role of the projection postulate. Therefore the name of L\"uders is used. On the other hand,
the paper \cite{EPR} played the crucial role in the nonlocal treatment of this postulate. Thus, {\it projection nonlocality can be called Einstein-L\"uders nonlocality.}

We start considerations with the extended citation from the practically unknown preprint of A. Aspect \cite{AA1}
(see also \cite{AA0}). In this preprint,  ``projection nonlocality''  came in all its brilliance.

\section{Alain Aspect: Counter-intuitiveness of quantum formalism}
\label{AAS}

``Let    us    consider    the    optical    variant    of    the    Bohm's    version    of    the    E.P.R.    Gedankenexperiment.   A   source   $S$ emits   a   pair   of   photons   with   different   frequencies $\nu_1$ and $\nu_2$, counterpropagating along $Oz.$ Suppose that the polarization part of the state vector describing the pair is: 
$$
\vert \Psi(\nu_1, \nu_2)\rangle= \frac{1}{\sqrt{2}} \{ \vert x,x\rangle + \vert y,y \rangle \},
$$                                                                             
where $x$  and  $y$  are  linear  polarizations  states.''

``Let us now consider the probabilities $p_{\pm \pm}(a,b)$ of joint detections of $\nu_1$ and $\nu_2$ in the  channels  +  or  - of  polarisers  $I$  or  $II,$  in  orientations  $a$   and  $b.$  Quantum  mechanics  predicts : 
$$
p_{++}(a,b)= p_{--}(a,b)=\frac{1}{2} \cos^2 (a,b), \; p_{+-}(a,b)= p_{-+}(a,b)=\frac{1}{2} \sin^2 (a,b)
$$
We  are  going  to  show  that  these  quantum  mechanical  predictions  have  far  reaching  consequences.''

\medskip

``As  a  naive  physicist,  I  like  to  raise  the  question  of  finding  a  simple  image  to  understand  these  strong  correlations.  The  most  natural  way  to  find  an  image  may  seem  to  follow  the  quantum  mechanical  calculations  leading  to  $p_{\pm, \pm}(a,b).$  In  fact,  there  are  several  ways  to  do  this  calculation.  A  very  direct  one  is  to  project  the  state  vector  
$\vert \Psi(\nu_1, \nu_2)\rangle$ onto  the  eigenvector  corresponding  to  the  relevant  result.  This  gives  immediately  the  joint  probabilities  $p_{\pm, \pm}(a,b).$  However, since this calculation bears on state vectors describing globally the two photons, I do not know how to build a picture in our ordinary space. 

In order to overcome this problem, and to identify separately the two measurements happening on both ends of the experiment, we can split the joint measurement in two steps. Suppose  for  instance  that  the  measurement  on  photon  $\nu_1$  takes  place  first,  and  gives  the  result + , with the polarizer $I$ in orientation $a.$ The + result (associated with the polarization state $\vert a\rangle)$ has a probability of 1/2. To proceed with the calculation, we must then use the postulate of reduction of the state vector, which states that after this measurement, the new state  vector  $\vert \Psi^\prime(\nu_1, \nu_2)\rangle$  describing  the  pair  is  obtained  by  projection  of  the  initial  state  vector $\vert \Psi(\nu_1, \nu_2)\rangle$  (equation  1)  onto  the  eigenspace  associated  to  the  result  +  :  this  two  dimensional  eigenspace  has  a  basis  $\{ \vert a,x\rangle, \vert a,y\rangle\}.$  Using  the  corresponding  projector,  we  find after a little algebra     
$$
\vert \Psi^\prime(\nu_1, \nu_2)\rangle= \vert a,a\rangle.
$$                                           
This means that immediately after the first measurement, photon $\nu_1$ takes the polarization $a$  :  this  is  obvious  because  it  has  been  measured  with  a  polarizer  oriented  along  $a,$  and  the  result  +  has  been  found.  More  surprisingly,  the  distant  photon  $\nu_2$  ,  which  has  not  yet  interacted with any polarizer, has also been projected into the state $a$ with a well defined polarization, parallel to the one found for photon $\nu_1.$ This surprising conclusion however leads to the correct final result (3), since a straightforward application of Malus law shows that a subsequent measurement performed along $b$ on photon $\nu_2$ will lead to  
$$
p_{++}(a,b)= \frac{1}{2} \cos^2 (a,b).                                          (8)
$$
The calculation in two steps therefore gives the same result as the direct calculation. But in addition it suggests a picture for the two steps measurement: 
\begin{itemize}
\item i: Photon $\nu_1$, which had not a well defined polarization before its measurement, takes the polarization associated to the obtained result, at the moment of its measurement: this is not surprising. 
\item ii: When  the  measurement  on  $\nu_1$  is  done,  photon  $\nu_2,$  which  had  not  a  well  defined  polarization before this measurement, is projected into a state of polarization parallel to the result of the measurement on $\nu_1.$  This is very surprising, because this change in the description  of  $\nu_2$  happens  instantaneously,  whatever  the  distance  between  $\nu_1$  and  $\nu_2$  at  the moment of the first measurement.
\end{itemize}

 This picture seems in contradiction with relativity. According to Einstein, what happens in a  given  region  of  space-time  cannot  be  influenced  by  an  event  happening  in  a  region  of  space-time that is separated by a space like interval. It is therefore not unreasonable to try to find  more  acceptable  pictures  for  ``understanding''  the  EPR  correlations.  It  is  such  a  picture that we consider now.''

The latter is the hidden variable picture  that led Bell to subquantum nonlocality which we call in his name {\it Bell nonlocality.}

Our aim is to show that the above picture does not contradict to relativity. Moreover, it is schematically  identical to the corresponding picture from classical probability theory, the picture of probability update - the basis of probability inference.

\section{ Einstein-L\"uders nonlocality}

In \cite{Luders},  L\"uders formalized in the form of a postulate the operation of the quantum state update resulting from measurement 
with the concrete output, as projection on the corresponding subspace of the state space. In fact, this postulate was actively used  from the first days of quantum theory, e.g., in the EPR-paper 
\cite{EPR}. Often misleadingly the projection postulate is associated with the name of von Neumann with referring to his book  \cite{VN}; often people even say about the von Neumann projection postulate (see footnote 1).  

\subsection{L\"uders update of quantum state}
\label{LPPPP}

In the original quantum formalism, an observable $A$ is represented  by a Hermitian operator acting in Hilbert state space $H,$
\begin{equation}
\label{T1}
 \hat A =\sum_x x\;  \hat E^A(x),
\end{equation}
where $E^A(x)$ is the orthogonal projector on the subspace $H_{x}$ composed of eigenvectors with  eigenvalue $x$
(we consider only observables with discrete spectra). 

For pure initial state $\vert \psi \rangle,$  the post-measurement state is always again the pure state given by 
normalized  projection: 
\begin{equation}
\label{T4}
\vert \psi_{A=x} \rangle =\hat E^A(x) \vert \psi\rangle/\Vert \hat E^A(x) \;\vert \psi \rangle \Vert.
\end{equation}
Thus measurement with output $A=x$ induces state update:
\begin{equation}
\label{Tm4}
\vert \psi\rangle  \to \vert \psi_{A=x} \rangle. 
\end{equation}
We also remind the Born rule for probability of $A$'s output: 
\begin{equation}
\label{T4m}
p(A=x; \psi) = \Vert \hat E^A(x) \;\vert \psi \rangle \Vert^2.
\end{equation}

From the first sight, the projection transformation of the state given by (\ref{T4}) has nothing to do with nonlocality
(neither the Born rule). However, by considering a compound system $S=(S_1,S_2)$ we shall obtain the state-update procedure making the impression of instantaneous action at a distance. 

Let quantum state $\vert \Psi\rangle$ belong to tensor product $H=H_1\otimes H_2$  of state spaces $H_i$ of systems $S_i, i=1,2.$ 
Select some observable $A$ on $S_1;$ it is represented by  Hermitian operator $\hat A$ given by (\ref{T1}), where $\hat E^A(x)$ 
acts in $H_1.$ This observable can be also treated as an observable on compound system $S.$ The latter is represented by projector
$\hat E^A(x)\otimes I$ in $H.$  By getting  output $A=x,$ we transform the state of $S$ on the basis of L\"uders projection postulate: 
\begin{equation}\label{A1}
\vert \Psi_{A=x}\rangle= \hat E^A(x) \otimes I \vert \Psi\rangle/\Vert \hat E^A(x) \otimes I \vert \Psi\rangle \Vert.
\end{equation}
Consider now an observable $B$ on $S_2$ and its conditional measurement, under output $A=x.$ By Born's rule    
\begin{equation}
\label{Ay}
p(B=y\vert A=x; \Psi)\equiv p(B=y; \Psi_{A=x}) =   \Vert \hat E^B(y)\otimes I \vert \Psi_{A=x}\rangle \Vert^2.
\end{equation}

If state $\vert \Psi\rangle $ is separable, i.e., $\vert \Psi\rangle = \vert \psi^{(1)}\rangle\otimes \vert \psi^{(2)}\rangle,$
then 
\begin{equation}
\label{Aym}
p(B=y\vert A=x; \Psi)= \frac{\Vert \hat E^B(y)\otimes \hat E^A(x) \vert \Psi\rangle \Vert^2}{ 
\Vert \hat E^A(x) \vert \psi^{(1)} \rangle \Vert^2} = \Vert \hat E^B(y) \vert \psi^{(2)}\rangle \Vert^2.
\end{equation}
Thus, for such a state, measurement of observable $A$ on $S_1$ does not change statistics for measurements of observable $B$ on 
$S_2,$
\begin{equation}
\label{Aymt}
p(B=y\vert A=x; \Psi) = p(B=y; \Psi)
\end{equation}

However, if a state is entangled, then generally
\begin{equation}
\label{Aymt1}
p(B=y\vert A=x; \Psi) \not= p(B=y; \Psi)
\end{equation}

We remark that any state of $S$ determines the states of  its subsystems $S_i, i=1,2;$ for the initial state $\vert \Psi\rangle,$ 
$$
\rho^{(1)}=\rm{Tr}_{H_2} \vert \Psi\rangle\langle \Psi \vert, \; 
\rho^{(2)}=\rm{Tr}_{H_1} \vert \Psi\rangle\langle \Psi \vert,
$$
and, for the post-measurement state  $\vert \Psi_{A=x}\rangle,$
$$
\rho_{A=x}^{(1)}=\rm{Tr}_{H_2} \vert \Psi_{A=x}\rangle\langle \Psi_{A=x} \vert, \; 
\rho_{A=x}^{(2)}=\rm{Tr}_{H_1} \vert \Psi_{A=x}\rangle\langle \Psi_{A=x} \vert.
$$  
The above considerations can be represented in the form of probabilities with respect to the states of $S_2.$
For separable state $\vert \Psi\rangle,$
\begin{equation}
\label{Aymt2}
\rho^{(2)}= \rho_{A=x}^{(2)} \;  \mbox{and} \; p(B=y; \rho^{(2)})= p(B=y; \rho_{A=x}^{(2)}),
\end{equation}
and, for entangled states, generally
\begin{equation}
\label{Aymt3}
\rho^{(2)} \not= \rho_{A=x}^{(2)} \; \mbox{and} \;  p(B=y; \rho^{(2)})\not= p(B=y; \rho_{A=x}^{(2)}),
\end{equation}

Above formulas are just mathematical expressions. To have some physical picture, we should present their interpretation.
The main issue (and problem) is state's interpretation.\footnote{ Nowadays, some experts in quantum foundations claim that all interpretations 
are equally useful; so it is meaningless to struggle for a ``right interpretation''; others claim that it is even possible to proceed 
without interpretation at all. I do not think so.}   

\subsection{Statistical vs. individual interpretations of a quantum state}

The statistical interpretation of a quantum state is commonly associated with the names of  
Einstein and Ballentine \cite{BL,BL1,BL2}.

\medskip

{\bf SI} {\it Quantum state $\vert \psi\rangle$ represents statistical features of a large ensemble of identically prepared quantum systems.}

\medskip
 
So, a quantum state is not the ``personal state'' of an individual quantum system, say of an electron.

The individual interpretation of a quantum state was originally used by the majority of the quantum community. It is very often even coupled to the  Copenhagen interpretation. However, we have to be careful by saying about the  Copenhagen interpretation. 
(Plotnitsky proposed to speak about interpretations in the spirit of Copenhagen \cite{PL1,PL2}.)
 Von Neumann definitely used {\bf II} \cite{VN}, but not Bohr \cite{BR0}. 

\medskip

{\bf II} {\it Quantum state $\vert \psi\rangle$ is the physical state of the concrete 
quantum system.} 

In particular, von Neumann considered the  Schr\"odinger equation as  describing the  dynamics of the physical state of a concrete quantum system, similarly to the Newtonian or Hamiltonian dynamics of a classical system.

\subsection{Individual interpretation: action at a distance for quantum states?}
\label{IINT}

For {\bf II},  the straightforward appeal to the  L\"uders projection postulate really makes the impression of  spooky action at a distance, so to say ``genuine quantum nonlocality''.  Let, for example, 
\begin{equation}\label{UAm}
\vert \Psi\rangle= (\vert 01\rangle + \vert 10\rangle)/\sqrt{2}, 
\end{equation}
where the vectors labeled as $\vert 0\rangle, \vert 1\rangle$ are eignevectors of $\hat A:H_1 \to H_1$ and $\hat B:H_2 \to H_2,$ 
and both $H_i$ are qubit spaces. (We omit the indexes for these vectors, i.e., it should be
$\vert 0\rangle_i, \vert 1\rangle_i, i=1,2.)$ 
 By measuring observable $A$ on $S_1$ and getting output $A=0,$ we found the compound system 
in the state $\vert \Psi_{A=0}\rangle= \vert 01\rangle.$ Thus the state of $S_2$ instanteneously becomes $\vert \phi\rangle=\vert 01\rangle.$  Since by {\bf II} a quantum state has the meaning of the physical state of the concrete system, this is nothing than action at a distance. 

\medskip

{\it By using {\bf II} for the quantum state, one really confronts with nonlocality, so to say genuine quantum nonlocality.}  

\medskip

However, this confrontation takes place only at the theoretical level, as can be seen in the next section. 

\subsection{Individual interpretation in lab}
\label{IIL}

It is the good place to stress that even those who use the individual interpretation of a quantum state understand well that all quantum predictions are of the statistical nature. One can speak   as much as he wants
about the wave function of the concrete electron, but in lab he would collect statistics.

For example,  von Neumann   \cite{VN} consistently used {\bf II}, but at the same time  he pointed that experimental verification is possible only through von Mises frequency approach to probability. In lab, he appealed  to {\bf SI}.  So, although by using {\bf II}  one confronts with action at a distance - at the level of the theoretical consideration, at the experimental level he is in the same situation as one using {\bf SI}.

Independently of an interpretation of the quantum state, the essence of the problem  is in the comment presented in the introduction.\footnote{As I know (from the private conversations), its author uses {\bf II}.} We are interested in its following part:

\medskip

{\it ``But the empirical fact remains, that setting a measurement operator on  site A  has an instantaneous ``influence'' o site B, if you later calculate conditional statistics between the data of site A and site B. And this faster-than-light change of statistical correlation is already empirically proven.''}

\section{Aspect versus EPR presentations}

The reader can see that in \cite{AA1} Aspect essentially followed the original EPR-reasoning \cite{EPR}. However, he excluded the most questionable component of the EPR-reasoning - the elements of reality.

The main problem of both presentations is the absence of the 
explicit statement on the interpretation of a quantum state, neither Einstein, Podolsky, and Rosen nor Aspect started with
its identification. This made their reasoning fuzzy and generated misinterpretations.\footnote{One of the problems of the quantum community is that ``one has to understood'' (typically from long conversations) what kind of state's interpretation is used by another. Debaters do not declare their interpretations from the very beginning. I can only dream for a conference, where everybody would have on his conference badge not only the affiliation, but also his interpretation of the wave function; say {\bf II} or {\bf SI}. It would be much easier to understand what people mean by their statements. If somebody thinks that he can proceed without assigning any interpretation to the wave function, then this should also be reflected on the badge. But, may be even the interpretation-badges would not help. Recently the reasoning  that projection$\implies$nonlocality was presented to me by one of the top experts in the many-worlds interpretation  of quantum mechanics.}   

Aspect uses {\bf II} (as I know from the private conversations) and by treating a quantum state as a state of an individual photon, he  confronts with projection 
nonlocality (section \ref{IINT}). However, this is mathematical nonlocality. In lab, everybody has to collect statistical data,
i.e., to use {\bf SI} (section \ref{IIL}). 

Now we recite partially  paper \cite{AA1}: ``This is very surprising, because this change in the description  of  $\nu_2$  happens  instantaneously,...'' So,  Aspect pointed to the instantaneous change of the description. But, why is this change in theoretical description  surprising? In fact, De Broglie \cite{DB} was surprised by such sort of surprising. 

The interpretational basis of this surprise  and the consequent  belief in quantum nonlocality is the mixture of the theoretical use of  {\bf II} in combination with the lab-use of  {\bf SI}. Reality of lab-collected statistical data makes the illusion of reality of a quantum state.  It seems that this quantum state reality fallacy has the origin in von Neumann book \cite{VN} (see section \ref{IIL}).

\medskip

We remark that {\it the EPR-paper was in fact directed against {\bf II}.} 
This is the good place to mention the Einstein-Bohr debate, \cite{EPR,BR}. It seems that 
Einstein emphasized troubles of quantum mechanics induced by {\bf II}. Bohr replied to  Einstein 
in the spirit of {\bf SI}. And by using the latter Bohr could not recognize the problem that was declared by Einstein:
incompleteness of quantum mechanics.

For our presentation, it is important  that in the EPR-paper measurements are conditional, first measurement on $S_1$ and then selection of some measurement on $S_2.$ So, corresponding probabilities are also conditional probabilities. Hence, by treating the EPR-Bohm 
probabilities as conditional, Aspect followed the EPR-paper. 

In fact, the conditional probability picture reflects properly the context of experiments testing violation of the Bell type inequalities. The joint measurement picture is used too straightforwardly. In real measurements, photons in lab 1 and lab 2 are not detected simultaneously. (Therefore, the time window should be introduced.) Thus, this is really the conditional measurement, first  for photon $\nu_1$ and then for photon $\nu_2,$ or wise verse. 

Moreover, even theoretically measurements in the EPR-Bohm experiments cannot be treated as joint measurements. Consider  
the von Neumann scheme \cite{VN} for joint measurement of two compatible observables, say $A$ and $B,$ 
with Hermitian operators $\hat A$ and $\hat B.$
To measure $A$ jointly with $B,$ one has to represent them as functions of the same observable, say $C,$ described  by operator $\hat C$ with nondegenerate spectrum; in terms of operators 
$\hat A= f(\hat C), \hat B= g(\hat C).$ But, in the EPR-Bohm framework this observable $C$ is nonlocal - in the usual sense, its measurement involves measurements in both labs (see \cite{KHRL} for discussion; the same viewpoint was presented in \cite{FS1,FS2}).       

\section{Excurse through classical probability update}
\label{CP}

Consider now classical probability theory. Probability space (Kolmogorov, 1933 \cite{K}) is a triple 
${\cal P}=(\Omega, {\cal F}, P),$ where $\Omega$ is a set of random parameters, ${\cal F}$ collection of subsets 
of $\Omega$ representing events, $P$ is a probability measure. 
\begin{itemize}
\item States of random systems are represented by  probability measures
\item Observables are defined as maps from $\Omega$ to $\mathbf{R}$ (having some special  property).
\end{itemize} 
Let  $A, B$ be  random variables (representing classical observables) with probability distribution 
$p_A(x), p_B(y)$ and conditional probability $p(B=y\vert A=x).$ We recall that the latter is given by the Bayes 
formula:
\begin{equation}\label{bb}
p(B=y\vert A=x)=P(B=y, A=x)/P(A=x), 
\end{equation}
for $P(A=x)>0.$ To make classical and quantum notations consistent, we 
set $p_A(x)= p(A=y; P),  p_B(y)= p(B=y; P)$ and $p(B=y\vert A=x)= P(B=y; P_{A=x}),$ where $P_{A=x}$ is the conditional probability corresponding to the output $A=x$ given by the Bayes formula: for an event $E,$  $P_{A=x}(E)= P(E\vert A=x)\equiv  P(E, A=x)/P(A=x).$

In accordance with Bayesian inference, by getting output $A=x$ of observable $A$ we perform update of the probability measure
(the direct analog of the quantum state update (\ref{Tm4})):
\begin{equation}\label{bb}
P \to  P_{A=x}.
\end{equation}
It induces the probability update
\begin{equation}\label{bb}
p(B=y; P) \to P(B=y; P_{A=x}).
\end{equation}

Now, consider a compound classical random system $S=(S_1, S_2).$ Let systems $S_i, i=1,2,$ are characterized by sets of random parameters 
$\Omega_i.$ Then $S$ is characterized by set $\Omega= \Omega_1 \times \Omega_2,$ Cartesian product. If random systems are {\it independent,}
then the distribution $P$ of random parameters of $S$ is given by $P= P^{(1)}\otimes P^{(2)},$ where $P_i, i=1,2,$ are
 probability measures for $S_i$ and tensor product of measures is defined as $P(E_1\times E_2)= P^{(1)}(E_1)P^{(2)}(E_2).$ 
We call such a probability state separable. Separability is just another term for independence of random variables $I_i: \Omega \to \Omega_i, I_i(\omega)= \omega_i.$  If a probability measure on $\Omega$ is not separable, we call it entangled. 

Let $A$ and $B$ be observables on random systems $S_1$ and $S_2$ given by random variables, $A: \Omega_1 \to \mathbf{R}$ 
and $B: \Omega_2 \to \mathbf{R}.$ They can be treated as observables on $S$ represented by random variables $A(\omega)= A(\omega_1),  
B(\omega)= B(\omega_2).$ It is easy to see that for a separable measure $P= P^{(1)} \otimes P^{(2)},$
\begin{equation}
\label{bbt}
P(B=y; P_{A=x}) =  p(B=y; P)
\end{equation}
cf. with the quantum case, (\ref{Aymt}).
So, the state update based on to output $A=x$ does not change probability distribution for $B$-observable. 
However, if $P$ is entangled, then generally the probability distribution of $B$ is modified:
\begin{equation}
\label{bbt1}
P(B=y; P_{A=x}) \not=  p(B=y; P), 
\end{equation}        
cf. with the quantum case, (\ref{Aymt1}).

We remark that, similarly to the quantum case, in the classical case state $P$ of $S$ determines states of $S_i,$ as the marginal probabilities:
\begin{equation}
\label{bbt3}
P^{(1)}(E_1)= \int_{E_1\times \Omega_2} d P(\omega_1, \omega_2), \;    
P^{(2)}(E_2)= \int_{\Omega_1 \times  E_2} d P(\omega_1, \omega_2), 
\end{equation}
for discrete parameter-spaces, integration is reduced to summation. If a state is separable, $P=P^{(1)}\otimes P^{(2)},$  then its marginals coincide with $P^{(1)},P^{(2)}.$ Hence, 
\begin{equation}
\label{bbt2}
P^{(2)}_{A=x}= P^{(2)}   \;  \mbox{and}  \;  P(B=y; P^{(2)}_{A=x}) =  p(B=y; P^{(2)}),  
\end{equation}
where $P^{(2)}_{A=x}$ is the marginal of the updated state $P_{A=x};$ cf. with the quantum case, (\ref{Aymt2}). For entangled state $P,$ this equality is violated. Generally
\begin{equation}
\label{bbt4}
P^{(2)}_{A=x} \not= P^{(2)} \; \mbox{and}   \;  P(B=y; P^{(2)}_{A=x}) \not=  p(B=y; P^{(2)}),
\end{equation}
cf. with the quantum case, (\ref{Aymt3}).

Thus by getting result $A=x,$ we ``instantaneously'' get to know that statistics of $B$-measurement is changed.
We can put this consideration into the spatial scenarios, say random variable $A$ is measured in V\"axj\"o and random variable $B$ in Moscow. By getting $A=x$ in V\"axj\"o one  immediately updates the probability to get the output $B=y$ in Moscow.
Thus, even in classical probability ``faster-than-light change of statistical correlation'' takes place. But, this is not surprising. Let us consider this 
situation from the viewpoint of two basic interpretations of classical probability\footnote{We remark that interpretations 
are important not only in quantum mechanics, but in probability theory as well.}:
\begin{itemize}
\item statistical, 
\item subjective
\end{itemize}
By the statistical interpretation, state update given by (\ref{bb}) means that in the original ensemble of systems with state (=probability distribution)
$P,$ observer selects the subensemble  with state (conditional probability distribution) $P_{A=x}.$ It is not surprising at all that this filtration can change statistics of outcomes for another observable $B.$ Of course, the physical filtration and creation of a new ensemble with probability measure
$P_{A=x}$ takes time. But, a theoretical prediction is (practically) instantaneous. It is based on correlations encoded in the original state $P.$ For subjectivists, the situation is even simpler, because probability is subjective and its update is a purely mental process. It happens (practically) instantaneously. (Here ``practically'' is mentioned, since even theoretical operations take 
some time.)

\section{V\"axj\"o interpretation: quantum theory as machinery for probability update}

{\bf SI} of a quantum state is based on the statistical interpretation of probability. 
Considerations of the quantum and classical updates of states and the corresponding probabilities (sections \ref{LPPPP}, \ref{CP}) are practically identical (up to symbols). Measurement of quantum observable $A$ with output $A=x$ leads to selection of a subensemble of quantum systems and it is natural that statistics of measurements of another observable $B$ is changed. Physical filtration 
w.r.t. the concrete output $A=x$ takes time, but the theoretical prediction is (practically) instantaneous.   

Typically, by referring to {\bf SI} of quantum states one emphasized the statistical probability interpretation of Born's rule
(\ref{T4m}). We equally emphasize the state modification resulting from the filtration w.r.t.  the concrete output $A=x,$  see eq.
(\ref{Tm4}). Thus, {\it quantum theory is treated as machinery for probability update.} This is the essence of {\it the  V\"axj\"o
interpretation} \cite{V2}. 

We also point to QBism, the subjective probability interpretation  of quantum mechanics. QBists' position is the same as classical subjectivists,  ``faster-than-light change of statistical correlation'' is a mental process (well, even this process takes some time) 
\cite{Fuchs6}. QBism also treats quantum theory is as machinery for probability update.\footnote{ However, the V\"axj\"o interpretation and QBism differs crucially not only by the interpretations of probability, statistical vs. subjective, but also by the rule of probability inference implied by the quantum state update  (\ref{Tm4}). In the V\"axj\"o approach, the classical formula of total probability is perturbed by the interference term. So, with decreasing of the term quantum inference is smoothly transformed into  classical inference.
QBism advertize a new formula that should play the role of the formula of total probability, it cannot be smoothly transferred into the 
classical one.}  

Up to now, we emphasized the similarity of classical and quantum schemes for probability update. 
The main difference between them is that the quantum formalism induces another calculus, it is based on state transformation via the L\"uders projection postulate. Thus, the whole story is not about physics,
but about the rules for update of the given probability distribution on the basis of measurement output. Where does this difference  come from? 

First of all, we remark that for compatible  observables quantum probability update coincides with the classical one. So, it is clear that the origin of the difference between classical and quantum probability is in the existence of incompatible observables.  In probabilistic terms (see \cite{KHR_CONT}),  {\it incompatibility of observables is equivalent to nonexistence of the joint probability distribution} (jpd). The mathematical formalism of quantum mechanics gives a new rule for the probability update based on the state update in complex Hilbert space,  this rule is more general than the classical one. Classical events form the {\it Boolean algebra} and quantum events form a {\it partial Boolean algebra} (a consistently coupled collection of Boolean algebras).

This is the good place to cite Feynman  \cite{FeynmanP}  (italic shrift was added by the author of this paper):

``From about the beginning of the twentieth century experimental physics
amassed an impressive array of strange phenomena which demonstrated the
inadequacy of classical physics. The attempts to discover a theoretical structure
for the new phenomena led at first to a confusion in which it appeared
that light,and electrons, sometimes behaved like waves and sometimes like
particles. This apparent inconsistency was completely resolved in 1926 and
1927 in the theory called quantum mechanics. The new theory asserts that
there are experiments for which the exact outcome is fundamentally unpredictable,
and that in these cases one has to be satisfied with computing
probabilities of various outcomes.   {\it But far more fundamental was the discovery
that in nature the laws of combining probabilities were not those of the
classical probability theory of Laplace.}''

The quantum calculus of probabilities for incompatible observables, i.e., those without jpd,  generates the EPR-Bohm(-Bell) correlations. Following Bell, it is commonly claimed that these correlations are nonlocal.
However, as was shown in \cite{NL}, the crucial issue is local incompatibility, e.g., incompatibility of spin projections 
for system $S_1$ (or $S_2).$ The EPR-Bohm(-Bell) correlations are local. However, since they are based on incompatible observables, they are nonclassical from the viewpoint of probability theory (they are not based on the common  jpd). The quantum calculus of probability 
is a special calculus of contextual probability, see \cite{} for the general framework. Here contextuality is understood as dependence 
of probability distribution on the experimental context. The crucial point is that (in accordance with Bohr's complementarity principle)
there exist incompatible experimental contexts that cannot be reduced to the common context.    

We conclude:

\medskip

{\it Einstein-L\"uders nonlocality,  based on the projection postulate, coincides with ``nonlocality'' of 
 classical update of probability. }

\medskip

Such ``nonlocality'' is fictitious. This is ``nonlocality'' of mental operations.     

\section{``Spooky prediction at a distance''?}
\label{SPD}

At the level of theory, probability update for spatially separated observables can be considered as a prediction at a distance. 
(But, we should not forget that this theoretical prediction is about the real physical process, filtration on the basis of measurement 
output $A=x.)$ Recently Plotnitsky argued \cite{PARX} that, for quantum observables, such a prediction at a distance is really spooky; his reply to Summhammer would be -  call this situation {\it ``spooky prediction at a distance''} (instead of spooky action at a distance).

The V\"axj\"o interpretation was elaborated for long ago \cite{V2} - the interpretation of quantum mechanics  as machinery 
for probability update. But, Plotnitsky invented very good notion ``prediction at a distance''. For compound systems, 
the V\"axj\"o interpretation \cite{V2} implies ``prediction at a distance''. 

For me, the main problem is that  Plotnitsky  called such prediction ``spooky''. 
By him, ``spookiness'' of quantum predictions is due to the impossibility to create any space-time picture explaining correlations. Personally, I am not happy, neither with spooky action nor  spooky prediction and in section \ref{HB} I shall explain 
my position.

\section{Hertz-Boltzmann viewpoint on creation of scientific theories}
\label{HB}

By criticizing Plotnitsky, I again (cf. \cite{KHRAN,KHRHERZ}) refer to the Hertz-Boltzmann \cite{HER,BZ1,BZ2,DA} methodology of scientific theories. By this methodology, there are two levels of scientific representation of natural phenomena:
\begin{itemize}
\item a theoretical model (TM)
\item an experimental model (EM)
\end{itemize}
TM respects the universal {\it law of cause and effect.}\footnote{It states that for every effect there is a definite cause, likewise for every cause, there is a definite effect.} EM provides consistent description and prediction for experimental data. Already 
in 19th century scientists  understood well (at least Hertz and following him Boltzmann) that experimental data is statistical and
its description and prediction are probabilistic. For them, it was clear that EM need not be causal
(cf. with von Neumann \cite{VN}, acausality of quantum measurements). Of course, TM and EM have to be coupled, TM$\to$EM.
However, coupling is not rigid, TM is not rigidly coupled to experiment, TM is our mental image (``Bild'') of  physical phenomena,
its main aim is to respect the law of cause and effect. In short, Hertz and Boltzmann by developing the ``Bild-concept'' 
were precursors  of Bell with his attempt to introduce hidden variables in quantum theory. The main difference between the approaches of 
Hertz-Boltzmann and Bell is that Bell proposed the  very rigid rule connecting TM and EM (in this case EM=QM). He wanted that TM would describe the concrete outputs of quantum measurements: 
$$
A = A(\lambda). 
$$
Here $\lambda,$ a hidden variable, is an element of TM, the right-hand $A$ is also in TM, but the left-hand $A$ is in EM.

It is interesting that Bell was well aware about the problem of coupling of TM and EM (=QM). He started his activity 
\cite{Bell2} with really strong critique of von Neumann's no-go theorem \cite{VN}. Bell criticized precisely too rigid coupling
TM$\to$EM in von Neumann's consideration. In fact, all no-go statements are just statements about selection of possible $TM$ for 
$QM$ and the correspondence rule, TM$\to$EM. Unfortunately, creators of QM were not aware  about the ``Bild-concept'' of Hertz-Boltzmann (or just ignored it? Schr\"odinger tried to appeal to it, but his message was completely ignored). 

Coming back to the Einstein-Bohr debate, we can say that Einstein said that QM is not TM, Bohr replied that he did not see a problem, since he knows that QM is EM. It seems that Bohr did not reject a possibility to construct a consistent TM for QM (treated as EM), but he would not accept the Hertz-Boltzmann-Schr\"odinger  viewpoint on the structure of a scientific theory.\footnote{This is the good question 
to philosophers of science: Did Bohr and Einstein (as well as say Heisenberg and von Neumann) know about the works of Hertz and Boltzmann\cite{HER,BZ1,BZ2,DA}? The situation is really strange. Everything happened nearby Germany, all could read in German, and Hertz and Boltzmann were really famous.} He considered such kind of activity as metaphysical and, hence, meaningless. In contrast, Einstein badly wanted TM for QM, but (as latter Bell) he wanted too much from the map TM$\to$EM. 
\footnote{As to one of possible TM for EM=QM, we can point to {\it prequantum classical statistical field theory} (PCSFT) \cite{Beyond}, the classical random field model. Coupling of TM=PCSFT with EM=QM is very simple, a quantum state (density operator) is identified with the covariance operator of a complex random field that is normalized by the field energy; a quantum observable (Hermitian operator) corresponds to a quadratic form of a field.}           

Finally, we remark that modern philosophers operate with the similar scheme of the two levels structure of scientific theories \cite{Harald1}:
\begin{itemize}
\item ontic; 
\item epistemic.
\end{itemize}

It is surprising that philosophers (who really  read a lot) are not aware about the works of Hertz and Boltzmann.  
However, this not the main problem with the ontic-epistemic approach. The main problem is that the ontic level represents reality ``as 
it is'' (when nobody makes measurements). For Hertz and Boltzmann, TM was not about reality as it is, but just its mental ``Bild'', consistent and respecting the law of cause and effect. 

Now, turn to Plotnitsky's spooky predictions. Generally, predictions of any EM are spooky, since it is not EM's aim to present the causal picture of physical phenomena. The latter is done by TM. (Once again, QM$\not=$TM, QM=EM.) So, I think that terminology ``spooky predictions'' is misleading. 

\section{Concluding remarks}
\label{SRT}

The aim of this paper is disillusion of projection based nonlocality (Einstein-L\"uders nonlocality). This sort of nonlocality can be considered as genuine quantum
nonlocality, in contrast to subquantum (Bell equality based) nonlocality. It is important to distinguish sharply 
these two nonlocalities. They are often mixed in heads of scientists advertizing ``quantum nonlocality.'' This two faced Janus is often 
seen as having just one face - quantum nonlocality.

This mental mixing is explainable by taking into account coupling between  Einstein-L\"uders and Bell nonlocalities. This coupling was excellently 
presented in Aspect's paper \cite{AA1}. Einstein-L\"uders nonlocality makes quantum theory so counter-intuitive that any common sensible scientist  would try to find a beyond-quantum explanation. 

Bell proposed a class of subquantum models known as models with hidden variables \cite{Bell1,Bell2}. For such models,  he derived the inequality and its violation was interpreted by him as the evidence of another 
sort of nonlocality (Bell nonlocality). 

Then it happened something really amazing: {\it Bell nonlocality was elevated to quantum physics
and also was treated as quantum nonlocality.} Two faced Janus of quantum nolocality was born. 

In \cite{NL}, it was shown that in the purely quantum  framework violation of the Bell  type inequalities is a consequence of local incompatibility of observables (e.g., observables 
$A_1$ and $A_2$ on system $S_1).$ Thus Bell nonlocality has nothing to do with quantum mechanics. This is a feature of one very special 
class of subquantum models considered by Bell \cite{Bell1,Bell2}.

Well, Aspect's presentation demonstrates that one may  consider Bell nonlocality as 
grown from Einstein-L\"uders nonlocality (the use of the  projection postulate). So, he may found some trace of quantumness in Bell nonlocality, 
as grown from so to say the quantum seed - the projection postulate. Therefore it is important to destruct this seed. 

 In the present paper, it was shown that  Einstein-L\"uders nonlocality is the typical ``nonlocality'' of probability update,  similar to ``nonlocality'' of classical probability inference. Both faces of nonlocality-Janus were destructed.

In this paper, we emphasized the role of  two basic interpretations of a quantum state, individual (physical) vs. statistical.
Following Aspect's reasoning and {\bf II}, one can really confront  Einstein-L\"uders nonlocality. On the other hand, {\bf SI} combined 
with treatment of quantum mechanics as machinery for probability update (the V\"axj\"o interpretation \cite{V2}) implies that Einstein-L\"uders nonlocality
is typical nonlocality of probability update, instantaneous modification of probability distribution. Instantaneous (up to the scale of brain's functioning) action takes place in mental and  not in physical space. 
 
Finally, coming back to Summhammer's comment cited in introduction and his question, I say that the right scientifically justified terminology for  ``faster-than-light change of statistical correlation''  is  {\it probability update on the basis of the quantum calculus.} We repeat that the latter is the probability calculus designed for operating with incompatible observables, i.e., those 
without jpd.


\begin{thebibliography}{999}

\bibitem{NL} A. Khrennikov, Get rid of nonlocality from quantum physics. Entropy, 21(8), 806 (2019).

\bibitem{ABELL} Khrennikov, A. After Bell. {\it Fortschritte der Physik (Progress in Physics)}.
Topical issue – International Conference Frontiers of Quantum and Mesoscopic Thermodynamics Prague, 
Czech Republic 27 July - 1 August 2015. {\bf 2017},  {\it 65},  N 6-8, 1600014.

\bibitem{ARV} A. Khrennikov, Quantum versus classical entanglement: eliminating the issue of quantum nonlocality.
 arXiv:1909.00267v1 [quant-ph].

\bibitem{AA0} A. Aspect, Experimental tests of Bell's inequalities in atomic physics, in Atomic Physics 8, 
Proceedings of the  Eighth  International  Conference  on  Atomic  Physics,  edited  by  I.  Lindgren,  A.  Rosen  and  
S.  Svanberg  (1982).

\bibitem{AA1} A. Aspect, Bell's Theorem: The naive view of an experimentalist. quant-ph/0402001.

\bibitem{Bell1} Bell, J.S. \textit{Speakable and Unspeakable in Quantum Mechanics}, 2nd ed.;  Cambridge University Press: Cambridge, UK, 2004.  

\bibitem{Bell2} Bell,  J.S.  On the problem of hidden variables in quantum theory. 
{\it Rev. Mod. Phys.} {\bf 1966}, {\it 38}, 450.

\bibitem{Luders} G. L\"uders, \"Uber die Zustands\"anderung durch den Messprozess. Ann. Phys. (Leipzig) 8, 322–328 (1951).

\bibitem{VN} Von Neumann, J. {\it Mathematical Foundations of Quantum Mechanics}; Princeton University Press: Princeton, NJ,
USA, 1955.

\bibitem{DV}   Davies, E. B.  {\it Quantum theory of open systems},  (Academic Press,  London, 1976).

\bibitem{Oz} Ozawa, M.  Quantum perfect correlations. {\it An. Phys.} {\bf 321},  744-769 (2006).

\bibitem{Oz1} Ozawa, M.  Probabilistic interpretation of quantum theory. {\it New Generation Computing} {\bf 34}, 125-152 (2016).
\bibitem{EPR} Einstein, A.; Podolsky, B.; Rosen, N.   Can quantum-mechanical description of physical reality be considered complete? \emph{Phys. Rev.} {\bf 1935}, {\it 47}, 777--780.

\bibitem{Muynck} De Muynck, W. {\it Foundations of Quantum Mechanics, an Empiricist Approach}; Springer: Dordrecht, 2006.

\bibitem{Boughn1} Boughn, S. Making sense of Bell's theorem and quantum nonlocality. {\it Found. Phys.} {\bf 2017}, {\it 47}, 640--657.

\bibitem{Griffiths} Griffiths, R.B.  Quantum nonlocality: Myth and reality. \emph{arXiv} \textbf{2019}, 
arXiv:1901.07050.

\bibitem{V2}  Khrennikov, A.  V\"axj\"o interpretation-2003: Realism of contexts. 
In  \emph{Quantum Theory: Reconsideration of Foundations}; Khrennikov A., Ed.; V\"axj\"o Univ. Press: V\"axj\"o, 2004, pp. 323--338.

 \bibitem{PARX} A. Plotnitsky,  ``Without in any way disturbing the system'': Illuminating the issue of quantum nonlocality.
 	arXiv:1912.03842 [quant-ph].

\bibitem{DB} De Broglie, L. (1964). The current interpretation of wave mechanics: a critical study. Elsevier.

\bibitem{BR} Bohr, N.  Can quantum-mechanical description of physical  reality be considered complete? \emph{Phys. Rev.} {\bf 1935},
 {\it 48},  696--702.
		
\bibitem{BL} Ballentine, L. E. (1989).  The statistical interpretation of quantum mechanics, \emph{Rev. Mod. Phys.},  42, 358--381.

\bibitem{BL1} Ballentine, L. E. (1998). \emph{Quantum Mechanics: A Modern Development}; WSP: Singapore.

\bibitem{BL2} Ballentine, L. E. (2001). Interpretations of probability and quantum theory, In:  Khrennikov, A. Yu. (ed)  \emph{Foundations of Probability and Physics, Quantum Probability and White Noise Analysis}, 13,  71--84, WSP: Singapore.
		
\bibitem{PL1} Plotnitsky, A.  \emph{Epistemology and Probability: Bohr, Heisenberg, Schr\"odinger and the Nature of   Quantum-Theoretical Thinking}; Springer: Berlin, Germany; New York, NY, USA, 2009.

\bibitem{PL2} Plotnitsky, A.  {\it Niels Bohr and Complementarity: An Introduction}; Springer: Berlin, Germany; New York, NY, USA, 2012.

\bibitem{BR0} Bohr, N.    \emph{The Philosophical Writings of Niels Bohr};  Ox Bow Press: Woodbridge, UK, 1987.

\bibitem{KHRL} A. Yu. Khrennikov, The role of von Neumann and Luders postu-lates in the Einstein, Podolsky, and Rosen considerations: Comparing measurements with degenerate and nondegenerate spectra. J. Math.Phys.,49, N 5, art. no. 052102 (2008).

\bibitem{FS1} Filipp, S.; Svozil, K. Tracing the bounds on Bell-type inequalities. AIP Conf. Proc. 2005, 750, 87–94.

\bibitem{FS2} Filipp, S.; Svozil, K. Generalizing Tsirelson’s bound on Bell inequalities using a min-max principle.
Phys. Rev. Lett. 2004, 93, 130407.

\bibitem{K} Kolmolgoroff, A.~N.  (1933). \emph{Grundbegriffe der Wahrscheinlichkeitsrechnung}, (Springer-Verlag, Berlin); 
Kolmolgorov, A.~N. (1936). \emph{The basic notions of probability theory}. 

\bibitem{Fuchs6} Fuchs, C.~A.,  Mermin, N.~D. and Schack, R. (2014). An Introduction to QBism with an Application to the 
Locality of Quantum Mechanics,  \emph{Am. J. Phys.}  82, p. 749.

\bibitem{KHR_CONT} Khrennikov,  A. (2009). \emph{Contextual Approach to Quantum Formalism}, (Springer, Berlin-Heidelberg-New York).

\bibitem{FeynmanP} Feynman, R. P.  (1951). \emph{The Concept of probability in quantum mechanics}; Berkeley Symp. on Math. Statist. and Prob. Proc. Second Berkeley Symp. on Math. Statist. and Prob. (Univ. of Calif. Press), 533--541.

\bibitem{KHRAN}  Khrennikov, A.  (2017). Quantum epistemology from subquantum ontology: Quantum mechanics from theory of classical random fields, \emph{Annals of Physics},  377,  147--163.

\bibitem{KHRHERZ} Khrennikov, A. Hertz's viewpoint on quantum theory. {\it Act Nerv Super} 61, 24 (2019). https://doi.org/10.1007/s41470-019-00052-1


\bibitem{HER} Hertz, H. (1899). The principles of mechanics: presented in a new form. London: Macmillan.

\bibitem{BZ1} Boltzmann, L. (1905). Uber die Frage nach der objektiven Existenz der Vorgnge in der unbelebten Natur. In Barth, J.A. (Ed.) Populre Schriften: Leipzig.

\bibitem{BZ2} Boltzmann, L. (1974). On the development of the methods of theoretical physics in recent times. In McGuinness, B. (Ed.) Theoretical physics and philosophical problems. Vienna circle collection, Vol. 5. Dordrecht: Springer.

\bibitem{DA} D'Agostino, S. (1992). Continuity and completeness in physical theory: Schr\"odinger's return to the wave interpretation of quantum mechanics in the 1950's. In Bitbol, M., and Darrigol, O. (Eds.) E. Schr\"odinger: philosophy and the birth of quantum mechanics (pp. 339–360). Editions Frontieres: Gif-sur-Yvette.

\bibitem{Beyond} Khrennikov, A. (2014). Beyond quantum. Singapore: Pan Stanford Publ.

\bibitem{Harald1} Atmanspacher, H.  Determinism is ontic, determinability is epistemic, in H. Atmanspacher and  R.~C. Bishop (eds.), \emph{Between Chance and Choice: Interdisciplinary  Perspectives on Determinism} (Imprint Academic, Thorverton UK), 2002, pp. 49--74. 


\end{thebibliography}
 \end{document}